\documentclass[a4paper,11pt]{article}

\usepackage[utf8]{inputenc}
\usepackage[T1]{fontenc}
\usepackage[a4paper,margin=2.15cm,footskip=24pt]{geometry}
\usepackage{graphicx}
\usepackage{amsmath,amssymb,amsfonts,bm}
\usepackage{booktabs}
\usepackage{array}
\usepackage{siunitx}
\usepackage{subcaption}
\usepackage[numbers,sort&compress]{natbib}
\usepackage{microtype}
\usepackage{xcolor}
\usepackage[normalem]{ulem}
\usepackage{hyperref}
\usepackage{placeins}
\usepackage{orcidlink}
\usepackage{fancyhdr}

\graphicspath{{figures/}}
\hypersetup{colorlinks=true, linkcolor=blue, citecolor=blue, urlcolor=blue}
\newcommand{\DeltaTA}{\Delta}
\newcommand{\eps}{\epsilon}
\newcommand{\dd}{\mathrm{d}}

\newcommand{\publicationnote}{%
  \footnotesize\itshape Accepted for publication in \textit{The European Physical Journal Plus}.%
}

\fancypagestyle{plain}{%
  \fancyhf{}
  \fancyhead[R]{\small\thepage}
  \fancyfoot[C]{\publicationnote}

}

\begin{document}

\title{\textbf{Trace Anomaly and Effective Topological Sources in Neutron Stars}}

\author{%
\begin{minipage}{0.92\textwidth}
\centering
Federico Nola\,\orcidlink{0000-0001-9249-9547}\\[0.7em]
\small
$^{1}$Dipartimento di Matematica e Fisica, Università degli Studi della Campania
``Luigi Vanvitelli'', viale Abramo Lincoln 5, I-81100 Caserta, Italy\\[0.25em]
$^{2}$Istituto Nazionale di Fisica Nucleare, Sezione di Napoli,
Strada Comunale Cinthia, 80126 Napoli NA, Italy\\[0.25em]
$^{3}$Istituto Nazionale di Fisica Nucleare, Laboratori Nazionali di Frascati,
C.P. 13, 00044 Frascati, Italy\\[0.6em]
Corresponding author: \href{mailto:federico.nola@unicampania.it}
{federico.nola@unicampania.it}
\end{minipage}
}

\date{}

\maketitle
\begin{abstract}
This work investigates whether the trace anomaly can diagnose the stellar response to a topological scalar field in tensor multi-scalar gravity. Eleven cold tabulated equations of state (EoSs) were examined, with six selected after conservative GR stability and speed of sound checks over the tabulated density range. Their fiducial topological configurations were compared with the corresponding GR models under complementary matching prescriptions. After controlling for stellar mass and EoS dependence, the GR trace source strength \(S_T\) remained strongly correlated with the topological mass response, with a partial Spearman coefficient \(\rho=0.975\) within the sampled fiducial sequences. Near \(1.4\,M_\odot\), the topological configurations were systematically more compact, with reductions in the Jordan frame radius of \(6.7\)--\(8.7\%\) at fixed baryonic mass and \(0.87\)--\(1.29\,\mathrm{km}\) at fixed gravitational mass. These shifts are comparable to current uncertainties from the Neutron Star Interior Composition Explorer (NICER) and their observational impact depends on the source. Despite the global deformation, the interior effective source remained dominated by matter, with a median topological contribution of about \(0.8\%\). The GR matter trace therefore emerges as a useful diagnostic of the fiducial topological stellar response.
\end{abstract}



\medskip

\section{Introduction}
\label{sec:introduction}

Neutron stars offer a unique probe of matter above nuclear saturation density, where the EoS determines key stellar observables \citep{LattimerPrakash2007,Oertel2017}. Multimessenger observations, particularly GW170817 and NICER mass--radius measurements, have provided increasingly stringent constraints on dense matter physics and neutron star structure \citep{Abbott2017GW170817,Abbott2018RadiiEOS,Miller2019NICER,Riley2019NICER,Miller2021J0740,Riley2021J0740}. Recent NICER analyses of PSR J0614--3329, PSR J0740+6620, and PSR J0437--4715 provide mass--radius constraints that allow a direct observational consistency test of the stellar deformations considered here \citep{Mauviard2025J0614,Salmi2024J0740,Miller2026J0437PublishedV5}.

To connect these observational constraints with the underlying dense matter properties, I characterized stellar matter using the normalized trace anomaly, a dimensionless measure of the departure from conformal behavior in dense matter \citep{Fujimoto2022,CollinsDuncanJoglekar1977},
\begin{equation}
    \DeltaTA
    =
    \frac{\eps-3P}{3\eps}
    =
    \frac13-\frac{P}{\eps},
    \label{eq:delta}
\end{equation}
where the conformal limit corresponds to \(\DeltaTA=0\). Its relation to the sound speed,
\begin{equation}
    c_s^2
    =
    \frac{\dd P}{\dd\eps}
    =
    \frac13-\DeltaTA-\frac{\dd\DeltaTA}{\dd\ln\eps},
    \label{eq:cs2_delta}
\end{equation}
shows how both the magnitude and density dependence of \(\DeltaTA\) encode the stiffness of the EoS \citep{Fujimoto2022,Jimenez2024,GaribayEckerRezzolla2026}. As an external consistency check, I compared the reconstructed trace anomaly profiles\footnote{\label{fn:semiclassical-trace-anomaly}
Trace anomaly effects have also been studied in black holes and compact star configurations in semiclassical gravity
\citep{CaiCaoOhta2010,Cai2014ConformalAnomaly,
CalzaCasalinoSebastiani2022,ElHanafyNashed2025}.} with the quasi-universal relations proposed by Ren and Lin \citep{RenLin2026TraceAnomaly}.

The modified gravity analysis was based on the topological neutron star solutions of tensor multi-scalar gravity, where the scalar fields map spacetime onto a nontrivial target space and support configurations with nonzero topological charge but vanishing asymptotic scalar charge \citep{DonevaYazadjiev2020Topological}. Their branch structure and stability have been investigated previously, together with extensions to slowly rotating stars \citep{DonevaYazadjievKokkotas2020,Danchev2020}, within the broader context of scalarization in strong gravitational fields \citep{DamourEspositoFarese1993,DamourEspositoFarese1996}.

The relevance of the matter trace to scalar extensions of gravity has also been emphasized in neutron star studies, where the sign of \(T^\mu{}_{\mu}\) can influence scalar sector phenomenology \citep{PodkowkaMendesPoisson2018}. Here, I instead test whether an integrated GR trace source can diagnose the response of topological tensor multi-scalar configurations.

Section~\ref{sec:framework} presents the theoretical framework, Sec.~\ref{sec:eos_methods} describes the EoS selection and numerical construction, Sec.~\ref{sec:results} reports the results, and Sec.~\ref{sec:discussion_conclusions} discusses their physical implications.

\section{Framework}
\label{sec:framework}

Within the perfect fluid approximation, the stress-energy tensor takes the standard form
\begin{equation}
    T_{\mu\nu}
    =
    (\eps+P)u_\mu u_\nu
    +
    Pg_{\mu\nu},
\end{equation}
where \(u^\mu\) is the fluid four-velocity \citep{Wald1984}. With the metric
signature \((-,+,+,+)\), its trace becomes
\begin{equation}
    T^\mu{}_\mu=-\eps+3P.
\end{equation}
I therefore defined the positive matter trace variable as
\begin{equation}
    S\equiv\eps-3P=-T^\mu{}_\mu,
\end{equation}
which is the same combination entering the normalized trace anomaly in
Eq.~\eqref{eq:delta} \citep{Fujimoto2022}. The GR reference configurations were obtained by integrating the
Tolman--Oppenheimer--Volkoff equations
\citep{Tolman1939,OppenheimerVolkoff1939},
\begin{equation}
\begin{aligned}
    \frac{\dd m}{\dd r} &=4\pi r^2\eps,\\
    \frac{\dd P}{\dd r} &=
    -\frac{(\eps+P)(m+4\pi r^3P)}{r(r-2m)}.
\end{aligned}
\end{equation}

The topological configurations were described within tensor multi-scalar gravity. In the Einstein frame and in the absence of a scalar potential, the action was taken as \citep{DonevaYazadjiev2020Topological,DonevaYazadjievKokkotas2020,Danchev2020}
\begin{equation}
\begin{aligned}
    \mathcal{S} &=
    \frac{1}{16\pi G_*}
    \int \dd^4x\sqrt{-g}
    \left[
        R
        -2\gamma_{ab}(\varphi)g^{\mu\nu}
        \nabla_\mu\varphi^a\nabla_\nu\varphi^b
    \right]+\\
    &+S_m\!\left[A^2(\varphi)g_{\mu\nu},\Psi_m\right].
\end{aligned}
\label{eq:tmsg_action}
\end{equation}
Matter was coupled to the Jordan frame metric
\(\tilde g_{\mu\nu}=A^2(\varphi)g_{\mu\nu}\)
\citep{DamourEspositoFarese1992,DamourEspositoFarese1993,DamourEspositoFarese1996}. The target space was chosen to be \(\mathbb S^3\),
\begin{equation}
    \gamma_{ab}\dd\varphi^a\dd\varphi^b
    =
    a^2\left[
    \dd\chi^2+\sin^2\chi
    \left(\dd\Theta^2+\sin^2\Theta\,\dd\Phi^2\right)
    \right],
\end{equation}
with \(\chi=\chi(r)\), \(\Theta=\theta\), \(\Phi=\phi\), and
\begin{equation}
    A(\chi)=\exp(\beta\sin^2\chi),\qquad
    \alpha(\chi)=\beta\sin(2\chi).
\end{equation}

The tabulated EoS variables were interpreted as Jordan frame quantities \((\tilde\epsilon,\tilde P)\), while the stellar structure was written in terms of the Einstein frame metric
\begin{equation}
\begin{aligned}
    \dd s_E^2&=-e^{2\Gamma(r)}\dd t^2+f(r)^{-1}\dd r^2+r^2\dd\Omega^2,\\
    f(r)&=1-\frac{2m(r)}{r}=e^{-2\Lambda(r)}.
\end{aligned}
\label{eq:metric_topo_revision}
\end{equation}
Introducing \(\psi=\chi'\), the resulting system of stellar structure equations was integrated in the form \citep{DonevaYazadjiev2020Topological}
\begin{equation}
\begin{aligned}
 m' &=4\pi r^2A^4\tilde\epsilon
      +\frac{a^2}{2}r^2 f\psi^2+a^2\sin^2\chi,\\
 \Gamma'&=\frac{r}{2f}\left[
 8\pi A^4\tilde P+a^2 f\psi^2
 -\frac{2a^2\sin^2\chi}{r^2}+\frac{2m}{r^3}\right],\\
 \chi'&=\psi,\\
 \psi'&=\frac{1}{f}\left[\frac{2\sin\chi\cos\chi}{r^2}
 +\frac{4\pi}{a^2}A^4\alpha(\chi)(\tilde\epsilon-3\tilde P)\right]-\\
 &-\left(\Gamma'-\Lambda'+\frac{2}{r}\right)\psi,\\
 \tilde P'&=-(\tilde\epsilon+\tilde P)\left[\Gamma'+\alpha(\chi)\psi\right],
\end{aligned}
\label{eq:topological_structure_system}
\end{equation}
where
\begin{equation}
    \Lambda'=\frac{m'/r-m/r^2}{f}.
\end{equation}
In the adopted \(G_*=c=1\) normalization, the target space coordinates and \(a^2\) were dimensionless, whereas \(s=\chi'(0)\) had units of inverse length.

Regularity and topology imposed
\begin{equation}
\begin{aligned}
 m(0)&=0,\qquad \chi(0)=n_{\rm top}\pi,\\
 \chi'(0)&=s,\qquad \tilde P(0)=P_c,
\end{aligned}
\label{eq:center_bc_revision}
\end{equation}
with finite \(\Gamma(0)\) \citep{DonevaYazadjiev2020Topological}. I started the integration at \(r_0=10^{-5}\,\mathrm{km}\) from the regular expansion and located the numerical surface at
\begin{equation}
 P_{\rm surf}=1.001\,P_{\min}^{\rm EoS}.
\end{equation}
Outside the stellar surface, the matter variables were set to zero and the scalar and metric equations were integrated up to \(r_{\rm out}=800\,\mathrm{km}\). For \(n_{\rm top}=1\), the asymptotic condition \(\chi(\infty)=0\) was imposed. Since the massless exterior solution decayed as \(\chi\sim r^{-2}\) \citep{DonevaYazadjiev2020Topological}, the shooting condition at finite radius was written as
\begin{equation}
    \mathcal R_s=r_{\rm out}\,\chi'(r_{\rm out})+2\chi(r_{\rm out}),
    \qquad |\mathcal R_s|\leq2\times10^{-3},
    \label{eq:shooting_residual_revision}
\end{equation}
which provided the criterion for accepting the asymptotic solution.

The fiducial theory parameters were chosen as
\begin{equation}
 \beta=0.08,\qquad a^2=10^{-3},\qquad n_{\rm top}=1.
 \label{eq:fiducial_parameters_revision}
\end{equation}
This choice is consistent with the representative \(n_{\rm top}=1\) setup of Ref.~\citep{DonevaYazadjiev2020Topological}. Wider parameter scans were performed only as stress tests, while the strong correlation claim was restricted to this fiducial slice.

The stored trace source strength was constructed from the GR reference configuration as
\begin{equation}
 S_T=\frac{1}{M_{\rm GR}}
 \int_0^{R_{\rm GR}}|\epsilon(r)-3P(r)|\,\dd V_{\rm proper},
 \label{eq:ST_definition_revision}
\end{equation}
with
\begin{equation}
 \dd V_{\rm proper}=4\pi r^2\left(1-\frac{2m(r)}{r}\right)^{-1/2}\dd r.
 \label{eq:GR_proper_volume_revision}
\end{equation}
The quantity \(S_T\) was dimensionless, normalized by \(M_{\rm GR}\), and evaluated solely over the GR stellar interior, excluding both scalar contributions and the exterior tail. The corresponding energy-weighted trace anomaly was defined as
\begin{equation}
\langle\Delta\rangle_\epsilon=
 \frac{\int_0^{R_{\rm GR}}\epsilon(r)\Delta(r)\,\dd V_{\rm proper}}
 {\int_0^{R_{\rm GR}}\epsilon(r)\,\dd V_{\rm proper}}.
 \label{eq:Delta_energy_weighted_revision}
\end{equation}

For the Einstein frame source decomposition, the matter and topological energy density contributions were written as
\begin{equation}
\begin{aligned}
    \eps_{\rm matter}^E &= A^4\eps,\\
    \eps_{\rm topo,kin}^E &= \frac{a^2}{8\pi}f\psi^2,\\
    \eps_{\rm topo,ang}^E &= \frac{a^2}{4\pi}\frac{\sin^2\chi}{r^2},
\end{aligned}
\end{equation}
with
\begin{equation}
\begin{aligned}
    P_{r,{\rm topo}}^E
    &=
    \frac{a^2}{8\pi}f\psi^2
    -\frac{a^2}{4\pi}\frac{\sin^2\chi}{r^2},\\
    P_{t,{\rm topo}}^E
    &=
    -\frac{a^2}{8\pi}f\psi^2,
\end{aligned}
\end{equation}
which gave
\begin{equation}
\begin{aligned}
    S_{\rm matter}^E &= A^4(\eps-3P),\\
    S_{\rm topo}^E
    &=\eps_{\rm topo}^E-P_{r,{\rm topo}}^E-2P_{t,{\rm topo}}^E,\\
    S_{\rm total}^E&=S_{\rm matter}^E+S_{\rm topo}^E.
\end{aligned}
\label{eq:effective_source_budget}
\end{equation}
This decomposition referred exclusively to effective field sources and did not represent a decomposition into microscopic particle or interaction contributions to the EoS.

\section{Equations of state and numerical strategy}
\label{sec:eos_methods}

\begin{table*}[!t]
\small
\centering
\caption{EoS screening results for \(n_B\geq n_0=0.16\,\mathrm{fm}^{-3}\). The table compares the central density at the first stable GR mass maximum, \(n_{c,\max}^{\rm stable}\), with the density \(n_{\rm flag}\) at which \(c_s^2\) first leaves the interval \(0<c_s^2\leq1\); \(n_{\rm flag}>n_{c,\max}^{\rm stable}\) indicates that the flagged region is not reached by the stable GR sequence.}
\label{tab:eos_selection}

\begin{tabular}{lccclc}
\toprule
EoS &
\(M_{\max}^{\rm GR}/M_\odot\) &
\(n_{c,\max}^{\rm stable}\) [fm$^{-3}$] &
\(n_{\rm flag}\) [fm$^{-3}$] &
flag &
status \\
\midrule

BL chiral & 2.080 & 1.159 & 1.195 &
\(c_s^2>1\), max \(1.066\) & not selected \\

BSk22 & 2.265 & 0.973 & 1.094 &
\(c_s^2>1\), max \(1.225\) & not selected \\

BSk24 & 2.279 & 0.968 & 1.096 &
\(c_s^2>1\), max \(1.224\) & not selected \\

BSk26 & 2.171 & 1.131 & 0.984 &
\(c_s^2>1\), max \(1.325\) & not selected \\

DDME2 & 2.483 & 0.822 &
\(1.310\) &
terminal derivative sensitivity & not selected \\

\midrule

SLy4 & 2.050 & 1.193 & -- & none & selected \\
FSU2H & 2.377 & 0.806 & -- & none & selected \\
FSU2R & 2.048 & 0.939 & -- & none & selected \\
DD2 & 2.418 & 0.858 & -- & none & selected \\
NL3 & 2.774 & 0.666 & -- & none & selected \\
MUSES \(npe\mu\) & 2.037 & 1.020 & -- & none & selected \\

\bottomrule
\end{tabular}
\end{table*}

The analysis considered eleven cold tabulated EoSs: SLy4, FSU2H, FSU2R, DD2, DDME2, BSk22, BSk24, BSk26, NL3, BL chiral and a MUSES Calculation Engine \(npe\mu\) model \citep{CompOSEweb,CompOSE2013,CompOSE2022,MusesCalculationEngine,Pelicer2025MUSES,DouchinHaensel2001SLy4,TolosCentellesRamos2017,Typel2010,HempelSchaffnerBielich2010,Lalazissis1997NL3,GorielyChamelPearson2013,Pearson2018}\footnote{The MUSES \(npe\mu\) EoS adopted here was previously used in Ref.~\citep{Arias-Aragon:2026mwz}, which provides additional details on the corresponding EoS calculation.}. The core sound speed \(c_s^2\) was obtained from a shape-preserving interpolation of \(P(\epsilon)\) for \(n_B\geq n_0=0.16\,\mathrm{fm}^{-3}\). EoSs were selected when their GR sequences were stable, supported \(M_{\max}\geq2\,M_\odot\), and satisfied \(0<c_s^2\leq1\) throughout the full tabulated core density interval probed in the preliminary EoS screening. Table~\ref{tab:eos_selection} summarizes these consistency checks and the resulting EoS selection. 

The screening is deliberately conservative and is applied over the entire tabulated density range considered. The subsequent analysis shows that, for BL chiral, BSk22, and BSk24, the speed of sound exceeds the causal limit only at densities above those reached by the corresponding stable GR configurations. In contrast, BSk26 reaches \(c_s^2>1\) within its stable branch, with \(c_s^2\simeq1.13\) at the first stable mass maximum. DDME2 remains regular throughout the stable branch, while a small terminal excursion, reaching \(c_s^2\simeq-0.008\) for the reference interpolation, appears only near the upper end of the tabulated density range around \(n_B\simeq1.31\,\mathrm{fm}^{-3}\) and depends on the interpolation procedure and table resolution. Thus, except for BSk26, the flagged behavior lies outside the stable GR sequence, and the criterion is used only to define a conservative benchmark set for the topological analysis.

For each selected EoS, the fiducial topological branch was constructed from 36 logarithmically spaced central pressures, starting near \(0.85\,M_\odot\). The shooting parameter was sampled with 100 logarithmically spaced negative values over \(10^{-6}\leq |s|\leq8\,\mathrm{km}^{-1}\). Candidate brackets were identified from sign changes of the asymptotic residual using integrations to \(250\,\mathrm{km}\) and refined with Brent's method \citep{Brent1973} using relative and absolute tolerances of \(10^{-8}\). The resulting solutions were validated by reintegration to \(800\,\mathrm{km}\) with the SciPy RK45 solver \citep{Virtanen2020SciPy}, using relative and absolute ODE tolerances of \(10^{-6}\) and \(10^{-8}\) and maximum radial steps of \(0.05\,\mathrm{km}\) inside the star and \(0.5\,\mathrm{km}\) outside. Local continuation and secant refinement were used only to recover short unresolved gaps. The corresponding central pressure ranges are given in Table~\ref{tab:pc_ranges}.

\begin{table}[!t]
\small
\centering
\caption{Central pressure ranges of the reconstructed fiducial branches. \(N\) is the number of configurations obtained for each EoS from the initial logarithmic grid. The \(N\) values in Table~\ref{tab:response_correlation} refer to the subset entering the correlation analysis.}
\label{tab:pc_ranges}
\begin{tabular}{lccc}
\toprule
EoS & \(N\) &
\(P_{c,\min}\) [km\(^{-2}\)] &
\(P_{c,\max}\) [km\(^{-2}\)] \\
\midrule
SLy4             & 34 & \(3.615\times10^{-5}\) & \(9.055\times10^{-4}\) \\
FSU2H            & 35 & \(2.531\times10^{-5}\) & \(6.504\times10^{-4}\) \\
FSU2R            & 36 & \(2.659\times10^{-5}\) & \(6.152\times10^{-4}\) \\
DD2              & 32 & \(2.541\times10^{-5}\) & \(6.229\times10^{-4}\) \\
NL3              & 34 & \(2.035\times10^{-5}\) & \(8.300\times10^{-4}\) \\
MUSES \(npe\mu\) & 36 & \(2.741\times10^{-5}\) & \(7.593\times10^{-4}\) \\
\bottomrule
\end{tabular}
\end{table}

Branch continuity was tracked through Hungarian matching \citep{Kuhn1955Hungarian} using
\[
D=d_s+0.35d_M+0.20d_R+0.15d_\chi,
\]
where \(d_s=|\ln|s_i|-\ln|s_j||\), \(d_M=|M_i-M_j|/(0.12\,M_\odot)\), \(d_R=|R_{J,i}-R_{J,j}|/(1\,\mathrm{km})\), and \(d_\chi=|\chi_{s,i}-\chi_{s,j}|/0.35\). Assignments with \(D>5\) were rejected. Only the connected branch below the first mass turning point was used in the physical analysis, following Ref.~\citep{DonevaYazadjievKokkotas2020}.

Numerical convergence was tested near the branch endpoints and first turning points. Tightening the ODE tolerances to \((10^{-8},10^{-10})\) and halving the radial steps changed the mass negligibly and the radius in the Jordan frame by at most \(0.412\%\). Increasing the exterior cutoff from \(800\) to \(1200\,\mathrm{km}\) changed the radius by at most \(0.149\%\), while local refinement shifted \(M_{\rm turn}\) by at most \(0.131\%\). The last three intervals before the turning point were further refined logarithmically, and doubling the shooting scan from 100 to 200 trial values produced changes below \(0.3\%\). The remaining small endpoint oscillations are therefore attributed to numerical sensitivity rather than to physical structure.

\begin{figure}[!t]
\centering
\includegraphics[width=0.75\columnwidth]{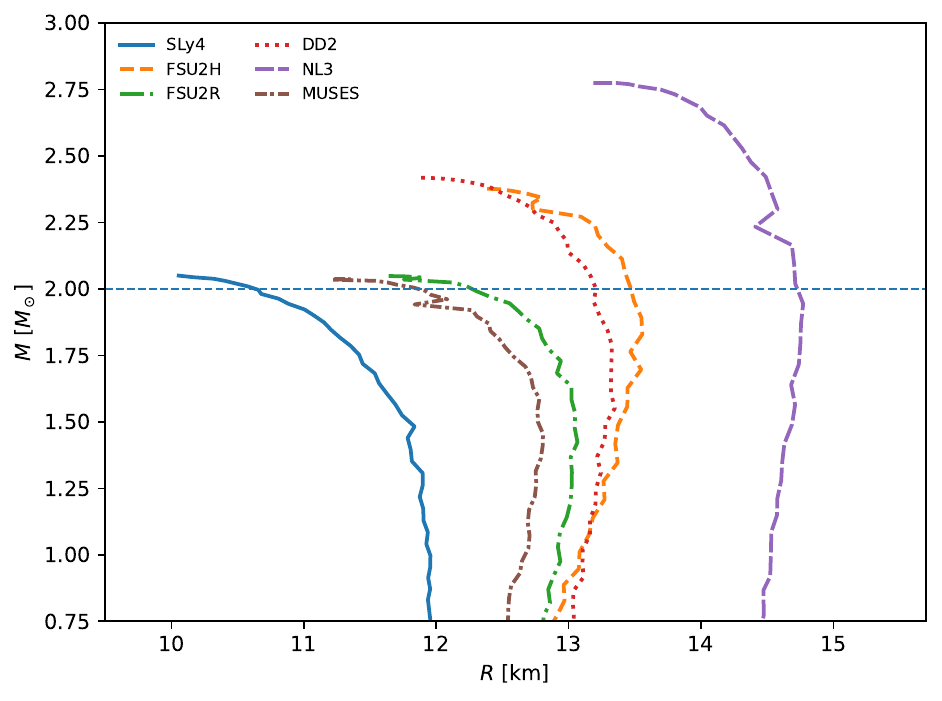}
\caption{Stable GR mass--radius sequences for the six selected EoSs, with the \(2\,M_\odot\) threshold indicated by the dashed blue line.}
\label{fig:mr_clean}
\end{figure}

\section{Results}
\label{sec:results}
\begin{figure*}[!t]
\centering
\includegraphics[width=1\textwidth]{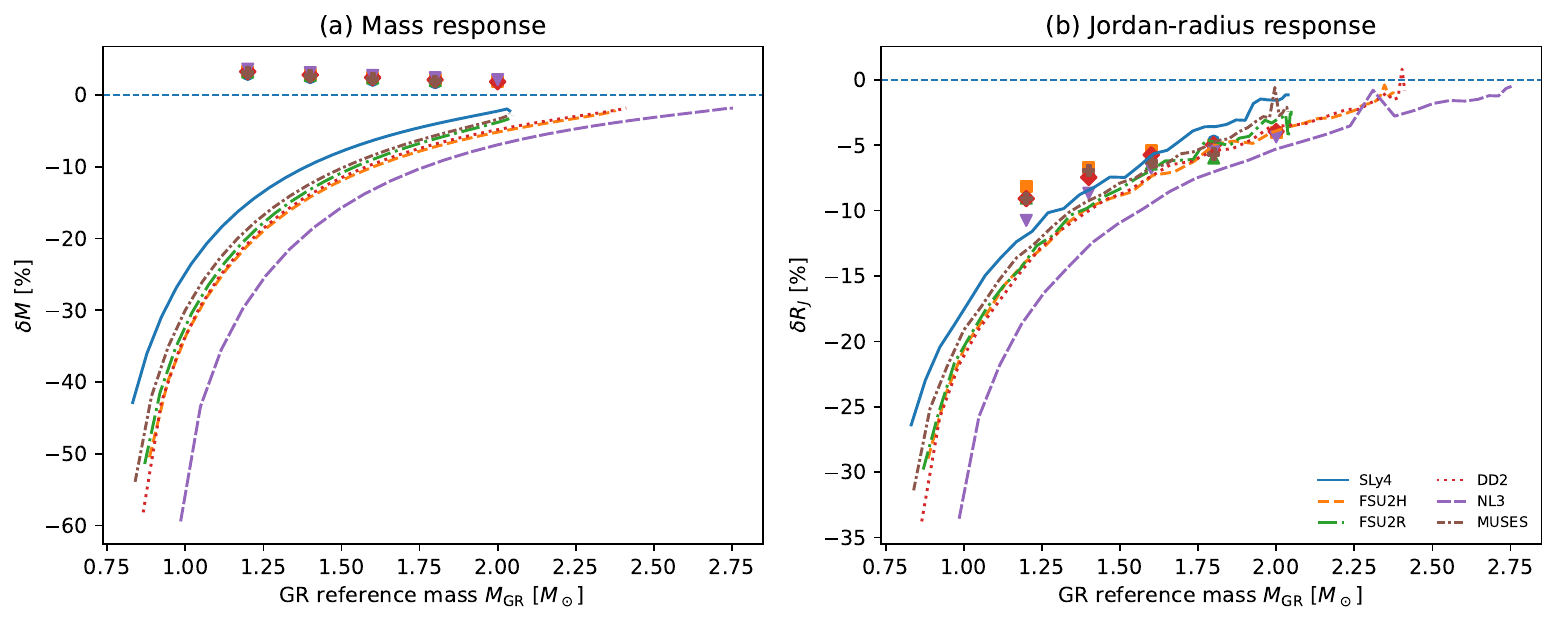}
\caption{Topological mass and Jordan frame radius shifts at fixed central pressure (curves) and fixed baryonic mass (symbols).}
\label{fig:fixed_pc_mb}
\end{figure*}

The GR sequences followed the expected stiffness ordering, with the central trace anomaly decreasing and the central sound speed increasing toward larger masses and compactnesses. Averaged over the selected EoSs, \(\Delta_c\) decreased from about \(0.23\) at \(1.2\,M_\odot\) to \(0.07\) at \(2.0\,M_\odot\), while \(c_{s,c}^2\) increased from about \(0.30\) to \(0.54\). The most compact SLy4 configuration reached \(\Delta_c\simeq-0.054\) and \(c_{s,c}^2\simeq0.79\), indicating that a negative central trace anomaly alone did not imply an EoS inconsistency.
 
The comparison with the topological configurations was performed under two
matching prescriptions. The first kept the EoS and central pressure fixed,
\begin{equation}
    P_c^{\rm topo}=P_c^{\rm GR}.
    \label{eq:fixed_pc_matching_revision}
\end{equation}
The resulting mass response is therefore defined at fixed \(P_c\), not at fixed gravitational mass. A GR reintegration at the same central pressure reproduced the reference masses to within \(1.91\times10^{-13}\).

For the topological configurations, the gravitational mass was evaluated as
\(M_{\rm topo}=m(r_{\rm out})\), while the physical circumferential radius was
taken in the Jordan frame,
\begin{equation}
    R_J=A(\chi_s)R_E,
    \label{eq:Jordan_radius_revision}
\end{equation}
leading to
\begin{equation}
\begin{aligned}
 \delta_M^{(P_c)}&=\frac{M_{\rm topo}(P_c)-M_{\rm GR}(P_c)}{M_{\rm GR}(P_c)},\\
 \delta_R^{(P_c)}&=\frac{R_J(P_c)-R_{\rm GR}(P_c)}{R_{\rm GR}(P_c)}.
\end{aligned}
\label{eq:delta_fixed_pc_revision}
\end{equation}

A second comparison was constructed at fixed conserved baryonic mass,
\begin{equation}
 M_b^{\rm GR}=4\pi m_b\int_0^{R_{\rm GR}}
 n_B(r)\frac{r^2\dd r}{\sqrt{1-2m(r)/r}},
 \label{eq:Mb_GR_revision}
\end{equation}
and
\begin{equation}
 M_b^{\rm topo}=4\pi m_b\int_0^{R_E}
 A^3(\chi)n_B(r)\frac{r^2\dd r}{\sqrt{1-2m(r)/r}},
 \label{eq:Mb_topo_revision}
\end{equation}
Here \(m_b\) denotes the atomic mass unit.
The matching was performed by interpolation along the connected branch below the first mass turning point. At fixed \(P_c\), the fiducial topological configurations were less massive and more compact than their GR counterparts, whereas at fixed \(M_b\) the radius remained smaller while the mass shift became positive. Near \(1.4\,M_\odot\), the latter increased the gravitational mass by \(2.43\)--\(3.13\%\) and reduced the Jordan frame radius by \(6.68\)--\(8.67\%\) (Table~\ref{tab:fixedMb14}). Fixed \(M_b\) results at \(2\,M_\odot\) were omitted when the corresponding baryonic mass lay outside the selected branch.

\begin{table*}[!t]
\small
\centering
\caption{Topological response near \(1.4\,M_\odot\) at fixed central pressure and fixed baryonic mass, with radii evaluated in the Jordan frame.}
\label{tab:fixedMb14}
\begin{tabular}{lrrrr}
\toprule
EoS & \(\delta_M^{(P_c)}\) [\%] & \(\delta_R^{(P_c)}\) [\%]
    & \(\delta_M^{(M_b)}\) [\%] & \(\delta_R^{(M_b)}\) [\%]\\
\midrule
SLy4 & -9.72 & -8.42 & +2.43 & -7.25\\
FSU2H & -14.22 & -9.74 & +2.87 & -6.68\\
FSU2R & -13.16 & -9.73 & +2.75 & -7.17\\
DD2 & -13.85 & -10.01 & +2.81 & -7.45\\
NL3 & -18.91 & -12.71 & +3.13 & -8.67\\
MUSES \(npe\mu\) & -12.38 & -9.24 & +2.66 & -6.93\\
\bottomrule
\end{tabular}
\end{table*}

\begin{figure}[!h]
\centering
\includegraphics[width=0.75\columnwidth,trim={0 0 0 20},clip]{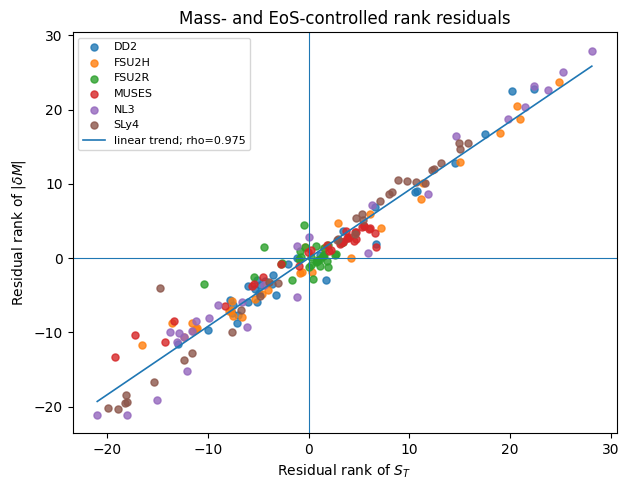}
\caption{Residual diagnostic for the partial Spearman analysis after removing the dependence on GR mass and EoS identity. The resulting correlation is \(\rho=0.975\).}
\label{fig:partial_rank_residuals}
\end{figure}

The correlation between \(S_T\) and the fixed-\(P_c\) mass response remained strong after controlling for stellar mass and EoS dependence. The individual EoS correlations and the combined and partial coefficients are reported in Table~\ref{tab:response_correlation}. The partial Spearman coefficient \citep{Spearman1904} was \(\rho=0.975\). It was obtained by rank transforming \(S_T\), \(|\delta_M^{(P_c)}|\), and \(M_{\rm GR}\), removing the dependence of the first two variables on ranked \(M_{\rm GR}\) and categorical EoS indicators by linear regression, and correlating the resulting residuals. Since the \(N=190\) configurations belong to six deterministic stellar sequences, this coefficient is interpreted as a descriptive measure of monotonic ordering rather than as a significance estimate for a statistical population. The result remained stable when each EoS was removed in turn, with \(\rho=0.972\)--\(0.975\) and when fewer configurations were sampled along each sequence, with \(\rho=0.930\)--\(0.975\). The corresponding residual diagnostic is shown in Fig.~\ref{fig:partial_rank_residuals}. Across the broader theory parameter scan, the correlation decreased to \(\rho\simeq0.516\), indicating that the strong ordering is specific to the fiducial theory slice.

\begin{table}[!t]
\small
\centering
\caption{Spearman coefficients between \(S_T\) and
\(|\delta_M^{(P_c)}|\) for each fiducial EoS and for the combined sample. The final row gives the partial coefficient after controlling for GR mass and EoS identity.}
\label{tab:response_correlation}
\begin{tabular}{lrr}
\toprule
Sample & \(N\) & \(\rho\)\\
\midrule
SLy4 & 33 & 0.998\\
FSU2H & 32 & 1.000\\
FSU2R & 33 & 1.000\\
DD2 & 31 & 1.000\\
NL3 & 28 & 1.000\\
MUSES \(npe\mu\) & 33 & 1.000\\
\midrule
All fiducial EoSs & 190 & 0.999\\
Partial, mass + EoS & 190 & 0.975\\
\bottomrule
\end{tabular}
\end{table}

\begin{figure}[!t]
\centering
\includegraphics[width=0.75\columnwidth]{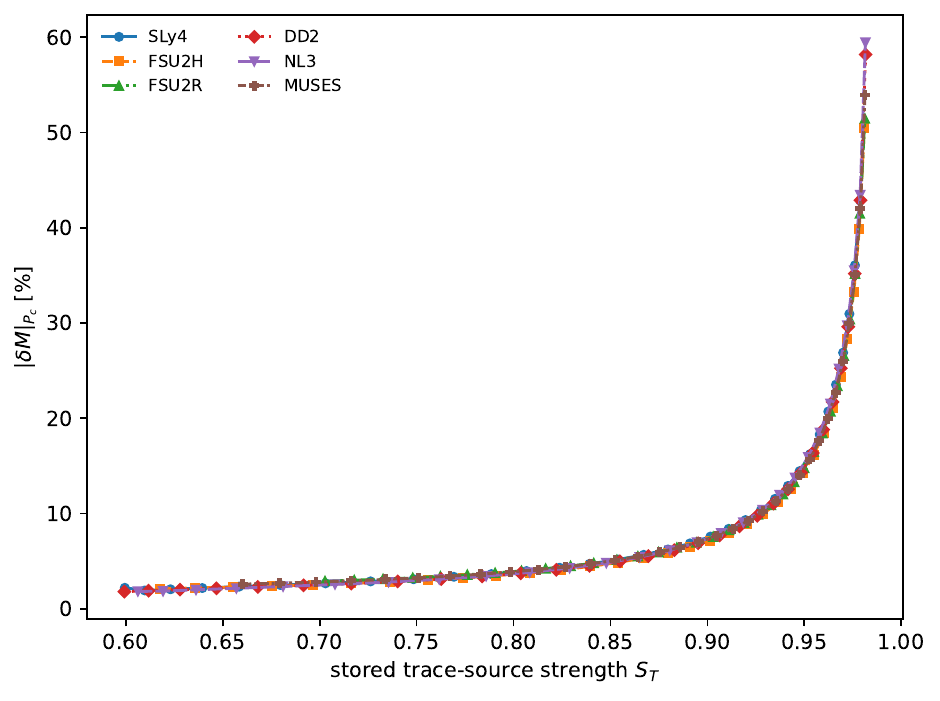}
\caption{Fixed \(P_c\) mass response versus the GR trace source strength \(S_T\) for the six fiducial EoSs.}
\label{fig:ST_response_revision}
\end{figure}

A complementary comparison was performed at fixed gravitational mass,
\begin{equation}
    \Delta R_J(M)=R_J^{\rm topo}(M)-R_{\rm GR}(M).
    \label{eq:deltaR_fixed_grav_mass}
\end{equation}
All six selected EoSs gave \(\Delta R_J<0\) over their common mass range. At \(1.4\,M_\odot\), the radius in the Jordan frame decreased by \(0.87\)--\(1.29\,\mathrm{km}\), corresponding to about \(7.2\)--\(8.9\%\). At \(2.0\,M_\odot\), only DD2, FSU2H, and NL3 remained available, with reductions of about \(0.5\)--\(0.7\,\mathrm{km}\). The correlation between \(S_T\) and the radius response at fixed mass was weaker, with \(\rho=-0.60\) at \(1.4\,M_\odot\) and \(\rho=-0.657\) at the J0614--3329 median mass. Given the limited sample of six EoSs, these values are treated as indicative trends rather than as evidence for a robust quasi-universal relation.

The mass--radius sequences were compared with public NICER posteriors for PSR J0614--3329, PSR J0740+6620, and PSR J0437--4715 \citep{Mauviard2025J0614,Salmi2024J0740,Miller2026J0437PublishedV5,Mauviard2025J0614DataV5,Salmi2024J0740DataV5,Miller2026J0437DataV5}. The topological compactification shifted the sequences toward the preferred radii of J0614--3329 and J0740+6620. For J0740+6620, the shift was \(0.47\)--\(0.62\,\mathrm{km}\), whereas the sequences moved away from the J0437--4715 posterior, which favors a larger radius. The observational effect therefore depends on the source.

\begin{figure*}[!t]
\centering
\begin{subfigure}{0.49\textwidth}
\centering
\includegraphics[width=\textwidth,trim=0 0 0 21px,clip]{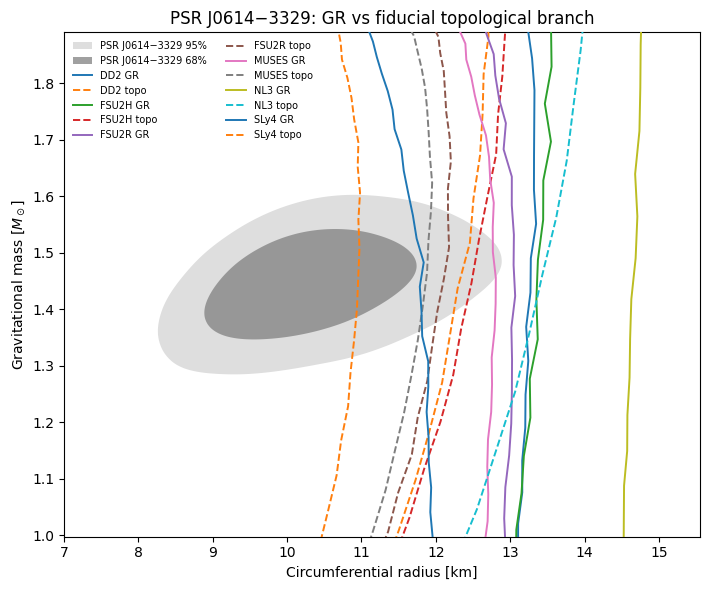}
\caption{J0614--3329 posterior with the GR and fiducial topological sequences.}
\end{subfigure}
\hfill
\begin{subfigure}{0.49\textwidth}
\centering
\includegraphics[width=\textwidth,trim=0 0 0 21px,clip]{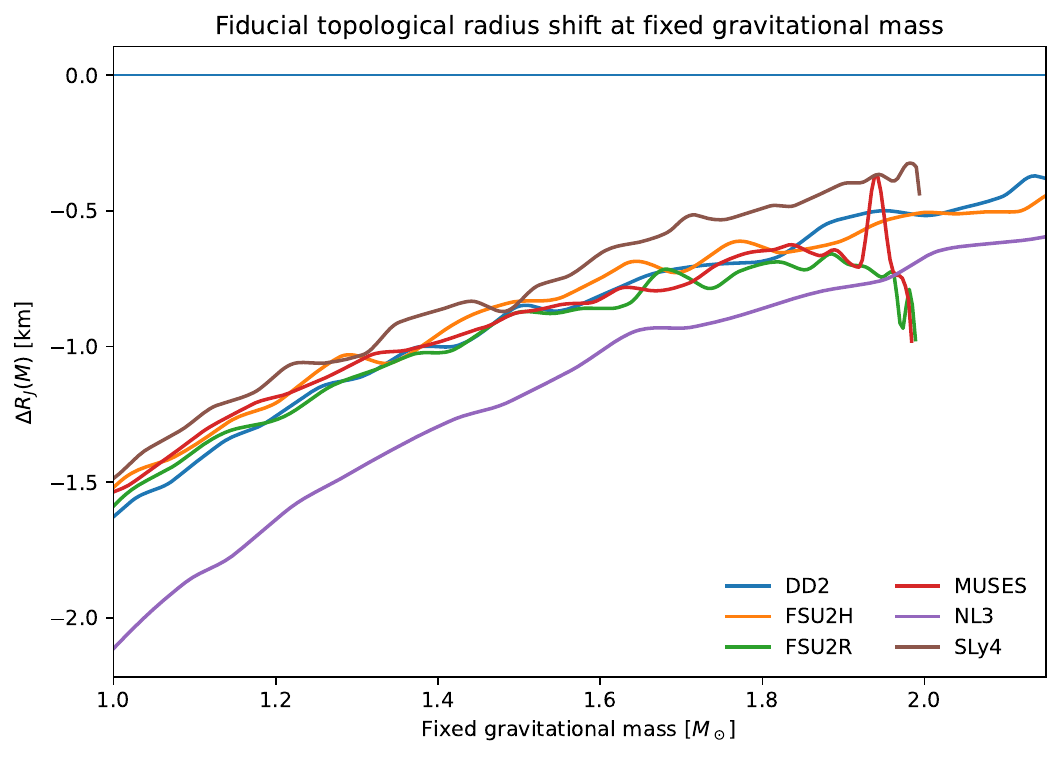}
\caption{Jordan frame radius shift at fixed gravitational mass.}
\end{subfigure}
\caption{Comparison with NICER constraints at fixed gravitational mass. Panel (a) shows the GR and topological sequences against the J0614--3329 posterior \citep{Mauviard2025J0614,Mauviard2025J0614DataV5}, while panel (b) shows the corresponding Jordan frame radius shift. Small variations close to the branch endpoints are attributed to the numerical sensitivity discussed in Sec.~3.}
\label{fig:nicer_radius_shift}
\end{figure*}

As an external consistency check, the GR trace anomaly profiles were compared with the Ren--Lin quasi-universal relations \citep{RenLin2026TraceAnomaly,RenLin2026Erratum}. At \(1.4\,M_\odot\), all but NL3 fell within the quoted \(\Delta_c\) range. The recent erratum provides polynomial coefficients with higher precision, which are required for an accurate numerical reconstruction of these relations, while leaving the main results and conclusions of Ref.~\citep{RenLin2026TraceAnomaly} unchanged. This correction does not affect the qualitative consistency check reported here. Accordingly, the Ren--Lin relations serve only as an external benchmark and do not enter either the topological reconstruction or the \(S_T\) correlation analysis.

The Einstein frame effective trace source was evaluated as
\begin{equation}
    S^E(r)=\epsilon^E(r)-P_r^E(r)-2P_t^E(r),
    \label{eq:trace_source_reduced}
\end{equation}
using the decomposition of Eq.~\eqref{eq:effective_source_budget}. The
integration was restricted to the stellar interior,
\begin{equation}
 \mathcal M_{\rm int}=\{r:\,P(r)>P_{\rm surf}\},
 \label{eq:Mint_revision}
\end{equation}
with
\begin{equation}
 \dd V_{\rm proper}=4\pi r^2\left(1-\frac{2m(r)}{r}\right)^{-1/2}\dd r.
\end{equation}

The relative topological contribution to the total trace source was quantified through
\begin{equation}
 \mathcal R_{\rm topo/total}^{\rm abs}=
 \frac{\int_{\mathcal M_{\rm int}}|S_{\rm topo}^E|\,\dd V_{\rm proper}}
 {\int_{\mathcal M_{\rm int}}|S_{\rm total}^E|\,\dd V_{\rm proper}},
\end{equation}
together with the bounded measure
\begin{equation}
 F_{\rm topo}^{\rm bounded}=
 \frac{\int_{\mathcal M_{\rm int}}|S_{\rm topo}^E|\,\dd V_{\rm proper}}
 {\int_{\mathcal M_{\rm int}}\left(|S_{\rm matter}^E|+|S_{\rm topo}^E|\right)\dd V_{\rm proper}},
 \label{eq:Fbounded_revision}
\end{equation}
and the signed and absolute source integrals
\begin{equation}
 Q_i=\int_{\mathcal M_{\rm int}}S_i^E\,\dd V_{\rm proper},\qquad
 Q_i^{\rm abs}=\int_{\mathcal M_{\rm int}}|S_i^E|\,\dd V_{\rm proper}.
 \label{eq:Qsigned_revision}
\end{equation}
Possible cancellations were quantified through
\begin{equation}
 C_{\rm canc}=1-\frac{Q_{\rm total}^{\rm abs}}
 {Q_{\rm matter}^{\rm abs}+Q_{\rm topo}^{\rm abs}}.
\end{equation}

The two measures differed by less than \(1.3\times10^{-5}\), with
negligible cancellations (\(C_{\rm canc}\leq1.35\times10^{-3}\)).
The signed topological contribution remained positive, while the bounded contribution was typically below \(1\%\), with median values of \(F_{\rm topo}^{\rm bounded}\) between \(0.779\%\) and \(0.857\%\) across the six EoSs. Only the NL3 configuration at the lowest mass reached \(3.55\%\). This enhancement reflected the comparatively small matter trace budget at low masses rather than a change of branch and it fell below \(1\%\) at higher masses.

\begin{figure}[!t]
\centering
\includegraphics[width=0.75\columnwidth]{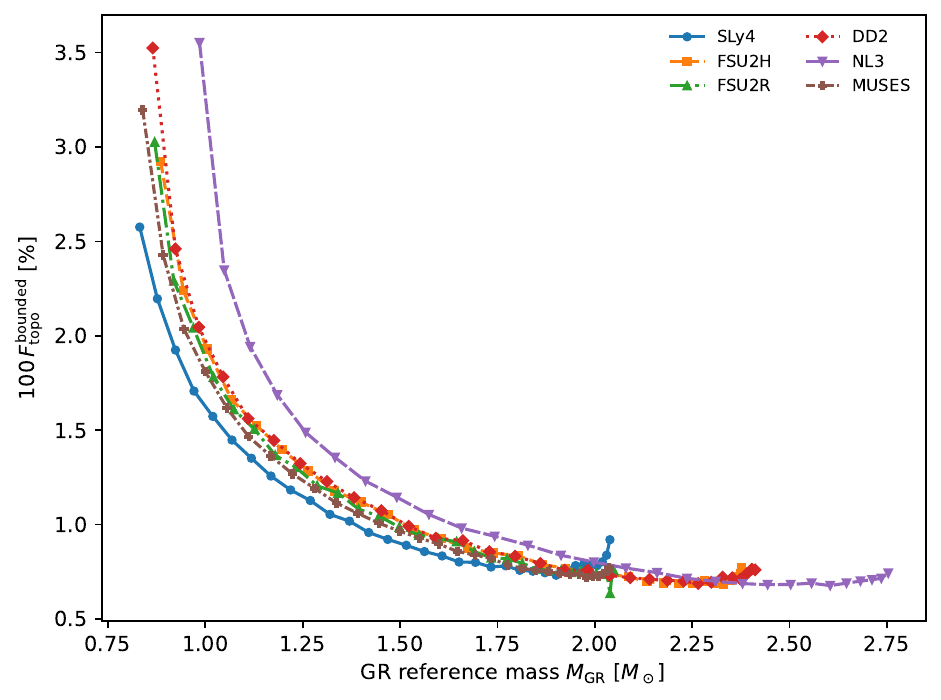}
\caption{Topological contribution to the interior trace source in the Einstein frame for the six fiducial EoSs. }
\label{fig:budget_integrated}
\end{figure}

\section{Discussion and conclusions}
\label{sec:discussion_conclusions}

The main result was that the GR trace source strength \(S_T\) provided a strong monotonic ordering of the fixed-\(P_c\) topological mass response within the fiducial theory slice, even after controlling for stellar mass and EoS dependence. This behavior weakened substantially in the broader parameter scan, indicating that the correlation was specific to the fiducial parameter choice rather than universal. At the same time, the bounded source budget in the Einstein frame remained strongly dominated by matter, with a median topological contribution of about \(0.8\%\), showing that a subdominant topological source can nevertheless accompany appreciable changes in the global stellar structure.

The physical interpretation depended on the matching prescription. At fixed \(P_c\), the fiducial topological configurations were both less massive and more compact than their GR counterparts, whereas at fixed baryonic mass they remained more compact but acquired \(2.4\)--\(3.1\%\) larger gravitational masses near \(1.4\,M_\odot\). At fixed gravitational mass, the corresponding radius reduction reached \(0.87\)--\(1.29\) km. These shifts are comparable to current NICER radius uncertainties and produced an observational effect that depended on the source: the topological sequences moved toward the preferred regions for J0614--3329 and J0740+6620, but away from the J0437--4715 posterior, which favors a larger radius. The comparison therefore indicates an observationally relevant deformation of the mass--radius relation, rather than evidence in favor of the topological sector itself.

Frame conventions and stability considerations are important when interpreting these results. The reported observables correspond to the asymptotic gravitational mass and the Jordan frame circumferential radius. The first mass turning points of the fiducial sequences ranged from \(1.984\) to \(2.703\,M_\odot\): DD2, FSU2H, and NL3 remained above the \(2\,M_\odot\) benchmark, whereas SLy4, FSU2R, and MUSES reached their first maximum slightly below it. The fiducial parameter choice should therefore be regarded as a controlled benchmark for the source analysis rather than as a generally viable point across all EoSs. Configurations below the first mass turning point were treated as consistent with stability following Ref.~\citep{DonevaYazadjievKokkotas2020}, although a dedicated analysis of radial modes would be required for a definitive stability assessment. Similarly, tidal deformability constraints require the corresponding perturbation equations in tensor multi-scalar gravity. Since the conformal factor approaches unity at large radii and the solutions carry vanishing asymptotic scalar charge \citep{DonevaYazadjiev2020Topological}, standard dipole radiation bounds derived for scalarized binaries with nonzero scalar charge cannot be applied directly and require a dedicated binary analysis.

Overall, the results suggest that the matter trace of the GR reference configuration can serve as a diagnostic of the global stellar response to the topological sector, despite its subdominant contribution to the integrated effective source. Future studies will be needed to determine how robust this correspondence remains across the theory parameter space and within more complete dynamical and observational treatments.

\FloatBarrier
\section*{Data availability}

The EoSs taken from CompOSE are publicly available at \url{https://compose.obspm.fr/}. The tables used in this work correspond to database records 134 (SLy4), 213 (FSU2H), 214 (FSU2R), 217 (DD2), 218 (DDME2), 253 (BSk24), 256 (BSk22), 258 (BSk26), 284 (NL3) and 306 (BL chiral). The MUSES \(npe\mu\) EoS adopted in this analysis was obtained with the MUSES Calculation Engine, available at \url{https://musesframework.io} and was previously used in Ref.~\citep{Arias-Aragon:2026mwz}, which provides additional details on the corresponding EoS calculation. The exact tabulated MUSES EoS used here is available from the corresponding author upon reasonable request. The NICER posterior datasets used in the calculations are publicly available through the permanent records cited in Refs.~\citep{Mauviard2025J0614DataV5,Salmi2024J0740DataV5,Miller2026J0437DataV5}.

\section*{Acknowledgments}

I thank Pietro Giarratana for introducing me to this topic and for the stimulating discussions that inspired the development of this work.

\section*{Author contributions}

The author was responsible for all aspects of this work.

\section*{Conflict of interest}

The author declares no conflict of interest.

\bibliographystyle{unsrtnat}

\bibliography{references_revision}

\end{document}